\documentclass{elsart}
\usepackage{amssymb}
\usepackage{amsmath}
\usepackage[dvips]{graphicx}

\journal{: IMA Volumes in Mathematics and its Applications, 52, pp. 111-121 (1993).\qquad\qquad\qquad\qquad\qquad\qquad}
\input{tcilatex}
\begin{document}

\begin{frontmatter}

\title{DYNAMIC EFFECTS IN GRADIENT THEORY FOR FLUID
MIXTURES}

\author{Henri Gouin}
\ead{henri.gouin@univ-cezanne.fr }

\address {Universit\'e d'Aix-Marseille \& C.N.R.S.  U.M.R.  6181, \\ Case 322, Av. Escadrille
 Normandie-Niemen, 13397 Marseille Cedex 20 France}

\begin{abstract}
 We propose a new method to study motions of mixtures in fluid interfaces.
 We extend the equations of equilibrium in interfaces and the results associated with traveling waves
for van der Waals like fluids [21]. Maxwell rule is extended to
interfaces of fluid mixtures
 out of equilibrium. Formulae like Clapeyron relation are obtained for isothermal layers.

\end{abstract}


\end{frontmatter}

{\small
\centerline{ (In: Shock Induced Transitions and Phase Structures in
General Media} \centerline{ Editors: R. Fosdick, E. Dunn and M. Slemrod)}}

\section{Introduction}

In this paper we propose a new attempt to study motions in interfaces of
fluid mixtures. First, we recall the study of interfaces between mixture
bulks in equilibrium. We remind problems for an interface moving in a single
fluid and extend them to cases of mixtures.

How to calculate the proportions of constituents in each bulk of a
n-components mixture? This problem is solved in the case of isothermal
equilibrium for mixtures whose free energy is known [7]. A simple calculus
consists in writing the total free energy of the mixture is minimum.

Practically, calculus are available only if:

- the container has a simple shape allowing one dimension calculus (cylinder
or sphere)

- the mixture has a few number of components

- the mixture free energy is not a too complex function of component
densities and temperature.

Let us denote by $\rho _i$ the density of constituent \emph{i}; $%
\displaystyle \rho = \sum_{i=1}^{n}\;\rho _ i$ is the density of the
mixture, $\displaystyle v = {\frac{1}{\rho}}$ the specific volume and $%
\displaystyle \ c_i = {\frac{\rho _ i }{\rho}}$ the concentration of
constituent \emph{i}. The volume free energy at temperature $\theta$ is in
the form $\,G_o = G_o(v, c_i, \theta).$ The total free energy is the sum of
the thermodynamic free energy and the interfacial energy.

The determination of the function $G_o$ is certainly of first importance,
but it is not the topic of this paper. This is both a theoretical and an
experimental problem and an equation of state with coefficients given by
experiments is used for calculations. The free energy must be independent of
the proportion of constituents when the density is vanishing [5,15,18].

In the case of flat interface, the knowledge of interfacial energy is not
necessary for the calculation of balance equations between the bulks. The
knowledge of interfacial energy is necessary to study spherical interfaces
(bulks, drops, aerosols...). A Landau-Ginzburg model consisting in a
quadratic form of density gradients representing the interfacial energy
[12]. Coefficients of the quadratic form are assumed to be constant (in the
case of a single fluid, this corresponds to a \textsl{mean field theory}
[16]). A quadratic form of the gradients of specific volumes can be also
used; nevertheless, assuming that the coefficients of the quadratic form are
constant, it has a different physical meaning. In fact, the choice of a
model is of no consequence outside of the interfacial layer. The
Landau-Ginzburg model represents an interface and the bulks and builds a
complete theory of the mixtures in dynamics [8,9,13].

In this paper, we apply the model to motions through interfaces for both
single fluids and mixtures. Because of physic scales, motions can be often
considered as permanent [4]. Interface moving in a single fluid was
investigated by Slemrod [20-21] as a traveling wave whose image in a
reference space is propagating with the constant velocity $C$. It is
equivalent to say the fluid crosses a stationary interface with the flow $q$
[4] ($q$ is the product of velocity and density). When dissipative phenomena
are neglected, the problem boils down to equilibrium with a specific free
energy increased with $\displaystyle {-{\frac{1}{2}}}\ q^2\ v^2$.

The general study of mixtures in dynamics can be applied in the case of
interfaces [9]. The constituent \emph{i} crosses the interface with the flow
$q_i $. We are back to a static problem with the same specific free energy
and an additional term in the form
\begin{equation*}
{-{\frac{1}{2}}}\ v^2\ \ \sum_{i=1}^{n} \ {\frac{{q_i}^2 }{c_i}} \;.
\end{equation*}
Each constituent of the mixture has its own reference space [1,9,17]. Due to
the fact that the interface has different velocities in each reference
space, the generalization of traveling wave is less intuitive.

For a $n$-component mixture, a permanent motion through an interface yields $%
n+1$ equations of \textit{dynamical} equilibrium. The equations represent a
linear system of the $n+1$ variables $q_i^2$ and $\theta$. When the specific
volume $v$ and the concentrations $\ c_i$ are given in each bulk, a simple
calculus yields the temperature of the mixture and the flow for each
constituent. Obviously, solutions are acceptable only if values $q_i^2$ and $%
\theta$ are positive.

Surprisingly, the problem of an interface in equilibrium is more difficult
to study than the dynamical problem. It is more difficult to choose $q_i^2$
and $\theta$ ($q_i$ is null in equilibrium case) and to deduce
concentrations and specific volume than to choose concentrations and
specific volume in each bulk and to deduce $q_i^2$ and $\theta$ from a
system of linear equations.

\section{Isothermal motion of a single fluid in the "Korteweg - van der
Waals" theory of capillarity.}

Let us consider the one dimensional motion of van der Waals like fluid. We
notice that the same equations describe a traveling wave of dynamic phase
transition and a flow through an interface in the case of permanent motion.

M. Slemrod [21] consider the one-dimensional motion of fluid possessing a
specific free energy of the form{\footnote{%
The notations are different from the ones given in [20-21]}:
\begin{equation*}
G(v,\theta )=G_{o}(v,\theta )+{\frac{e}{2}}({\frac{\partial v}{\partial X}}%
)^{2}
\end{equation*}%
where $X$ is the lagrangian mass variable, $e$ is a small positive
parameter. In the referential space, the balance laws of motion may be
written in the form:}

\begin{equation}
\left\{
\begin{array}{l}
\displaystyle{\frac{\partial v}{\partial t}}={\frac{\partial V}{\partial X}}
\\
\\
\displaystyle{\frac{\partial V}{\partial t}}={\frac{\partial }{\partial X}}\
\{-\mathit{p}(v,\theta )+\mu {\frac{\partial V}{\partial X}}-e^{2}{\frac{%
\partial ^{2}v}{\partial X^{2}}}\} \\
\  \\
\displaystyle{\frac{\partial E}{\partial t}}={\frac{\partial }{\partial X}}\
\{v\ (-\mathit{p}+\mu {\frac{\partial V}{\partial X}}-e^{2}{\frac{\partial
^{2}v}{\partial t^{2}}})+e^{2}\ ({\frac{\partial V}{\partial X}}{\frac{%
\partial v}{\partial X}})+k\ {\frac{\partial \theta }{\partial \xi }}\}%
\end{array}%
\right.
\end{equation}

where $V$ denotes the velocity of the fluid, $\mu $ the viscosity, $p$ the
pressure, $E$ the specific total energy, $k$ the coefficient proportionality
between the heat flux and the gradient of temperature in the Fourier Law and
$\xi =X-Ct$. The traveling wave theory of phase transitions attempts to
determine when two homogeneous phases may be joined by a traveling wave
solution $\{v=v(X-Ct)$, $V=V(X-Ct)$, $q=q(X-Ct)\}$ of equations (1). Slemrod
finds that the volume mass is solution of the equation:
\begin{equation*}
e^{2}{\frac{d^{2}v}{d\xi^{2} }}=-\,C^{2}(v-v_{o})-(p-p_{o})-\mu \,C\ v\eqno%
(2)
\end{equation*}%
$p_{o}$ and $v_{o}$ denotes the values of $p$ and $v$ in the bulks. Because
of the physical scales far from critical conditions, we established in [4]
the motions of van der Waals fluids crossing interlayers can be considered
as permanent. In the case of permanent motion of inviscid fluid, Eq. (1)$_2$
yields
\begin{equation*}
{\frac{dV}{dt}}={\frac{d}{dx}}\ \{-\mathit{p}(v,\theta )-e^{2}\ {\frac{%
\partial ^{2}v}{\partial X^{2}}}\}\;v
\end{equation*}%
where $x$ denotes eulerian variable and $\displaystyle{\frac{\partial x}{%
\partial X}}=v$ [20]. From $V=q\,v$ where $q$ denotes the constant flow of
the fluid for an unidimensional permanent motion, we deduce:
\begin{equation*}
e^{2}{\frac{\partial ^{2}v}{\partial X^{2}}}=-q^{2}(v-v_{o})-(p-p_{o})
\end{equation*}%
At $t$ given, the form of the equation is similar to Eq. (2), which allows
to obtain interpretations about dynamical Maxwell rule as in [20-21].

\section{Isothermal motion of fluid mixture in the "Korteweg - van der
Waals" theory of capillarity [9]}

For the sake of simplicity, we study a mixture of two fluids. The method can
be immediately extended to any number of constituents. No assumption has to
be done about composition or miscibility.

The motion of a two-fluid continuum can be represented with two surjective
differentiable mappings:
\begin{equation*}
z \rightarrow X_1 = M_1(z) \ \ \hbox{\rm and} \ \ z \rightarrow X_2 = M_2(z)
\end{equation*}
(Subscripts 1 and 2 are associated with each constituent of the mixture)

$X_1$ and $X_2$ denote the positions of each constituent in reference spaces
$D_{01}$ and $D_{02}$. The lagrangian of the mixture is:
\begin{equation*}
L = {\frac{1 }{2}} \ \rho _1 \mathbf{V}_1^2 + {\frac{1 }{2}}\ \rho_2 \mathbf{%
V}_2^2 - \varepsilon - \rho_1 \Omega _1 - \rho _2 \Omega _2
\end{equation*}
where $\mathbf{V}_1$ and $\mathbf{V}_2$ denote the velocity vectors of each
constituent, $\rho _1$ and $\rho _2$ are the densities, $\Omega _1$ and $%
\Omega _2 $ are the extraneous force potentials depending only on $z =(t,x)$
and $\varepsilon $ is the volume internal energy.

The expansion of the Lagrangian is in a general form. In fact dissipative
phenomena imply that $\mathbf{V}_1$ is almost equal to $\mathbf{V}_2$ and we
do not take into account some kinetic energy associated with the relative
velocity of the components. Because of the interaction between the
constituents, $\varepsilon$ does not divide into energies related to each
constituent of the mixture, like for \textsl{simple mixtures of fluids}
[14]. The mixture is supposed to be no chemically reacting. Conservations of
masses require:
\begin{equation*}
\rho _i \, \hbox {det}\ F_i = \rho _{oi} \ (X_i) \eqno (3)
\end{equation*}

where subscript i belongs to $\{$1,2$\}$. At t fixed, the Jacobian
associated with $M_i$ is denoted by $F_i$ and $\rho _{oi}$ is the reference
specific mass in $D_{oi}$.

In differentiable cases, Eq. (3) is equivalent to:
\begin{equation*}
{\frac{\partial \rho _i }{\partial t}} + \hbox{div} \ \rho _i \, \mathbf{V}%
_i = 0
\end{equation*}
The volumic internal energy $\varepsilon$ is given by the thermodynamic
behavior of the mixture. Each constituent has a specific mass; in the same
way, two specific entropies $s_1$ and $s_2$ are supposed to be associated
with constituents 1 and 2.

For an internal energy depending on gradients of densities, the volume
internal energy is:
\begin{equation*}
\varepsilon = \varepsilon (s_1 , s_2 , \rho _1 , \rho _2 , \hbox{grad} \rho
_1 , \hbox{grad} \rho _2)\eqno(4)
\end{equation*}
The quantity
\begin{equation*}
h_i = \varepsilon , _{\rho _i} - \varepsilon , _{\rho _{i ,\gamma , \gamma}}
\end{equation*}
\textsl{defines the specific enthalpy of the constituent i of the mixture}.
Subscript $\gamma$ corresponds to the spatial derivatives associated with
gradient terms; as usually, summation is made on repeated subscript $\gamma$%
.
\begin{equation*}
\theta _i = {\frac{\varepsilon ,_{ s_i} }{\rho _i}}
\end{equation*}
\textsl{defines the temperature of the constituent i of the mixture}.

In practise we use the expression
\begin{equation*}
\varepsilon = \alpha (s_1 , s_2 , \rho _1 , \rho _2 ) + {\frac{1 }{2}}\ Q
\end{equation*}

where $Q$ is a quadratic form with constant coefficients:
\begin{equation*}
Q = C_1 \,(\hbox{grad}\, \rho _1)^2 + 2D \,\hbox{grad} \rho _1 \,\hbox{grad}
\rho _2 + C_2\, (\hbox{grad}\,\rho _2 )^2
\end{equation*}
and $\alpha (s_1 , s_2 , \rho _1 , \rho _2 )$ is the value of internal
energy in the homogeneous bulks.

To obtain the equations of motions, we used variational principle whose
original feature is to choice variations in reference spaces (Fig. 1) [9].
\begin{figure}[h]
\begin{center}
\includegraphics[width=12cm]{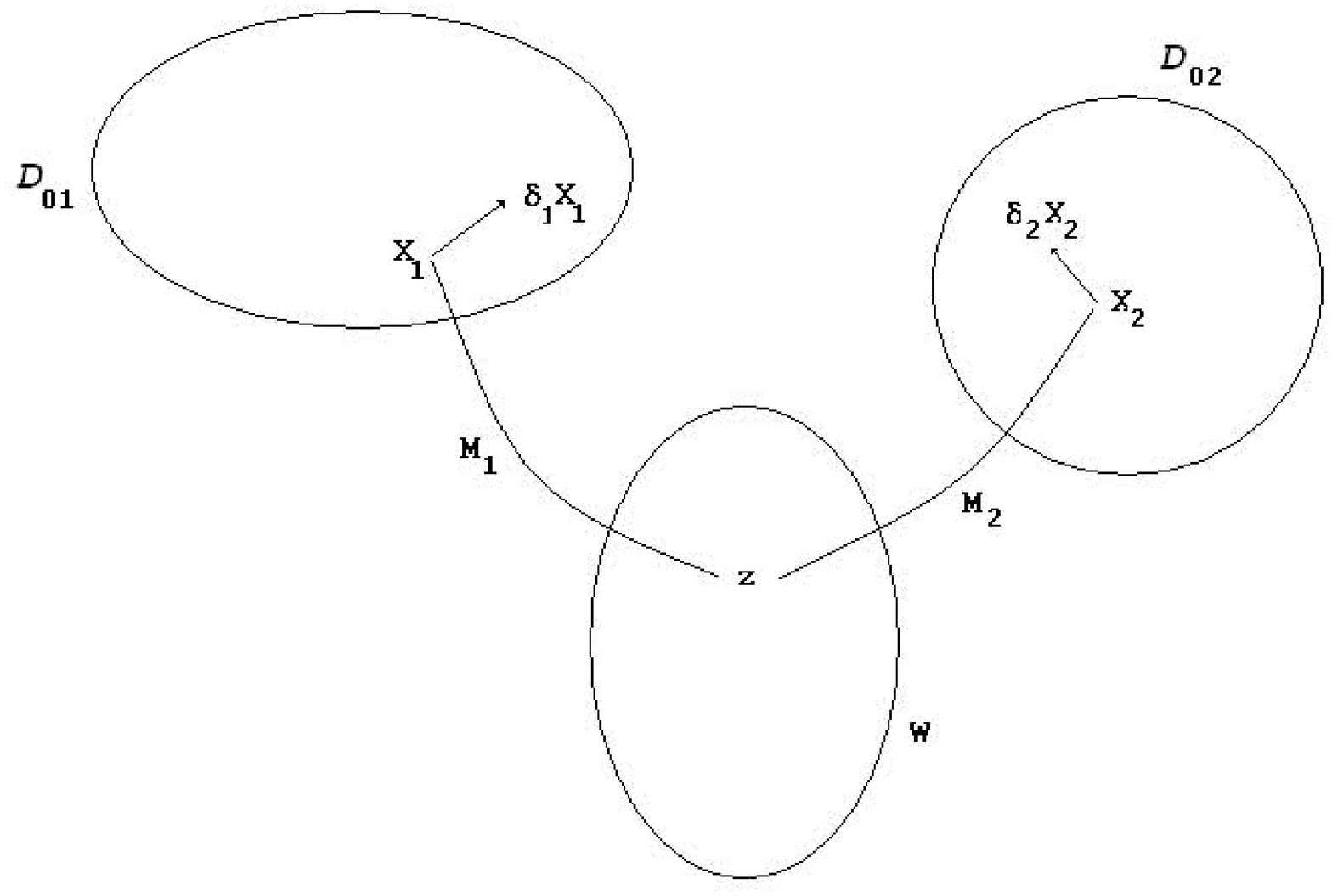}
\end{center}
\par
\label{Fig. 1}
\end{figure}

\vskip 0.25cm

\centerline{Figure 1}

Variations $\delta _1$ and $\delta _2$ of motions of particles are deduced
from $\ X_1 = \psi _1 (x, t, \beta _1)$ and $X_2 = \psi _2 (x, t, \beta _2)$%
. They are associated with a two-parameter family of virtual motions of the
mixture. The real motion corresponds to $\beta _1 = 0$ and $\beta _2 = 0$.
\newline
Obviously, a classical variation of function in space $W$ (as Serrin's p.
145, [19]) with only one parameter cannot give as many informations as the
ones given by any two-parameters family of virtual displacements.

The equation of the motion of each constituent of the mixture is:
\begin{equation*}
\Gamma _i = \theta _i \,\hbox {grad} \,s_i - \hbox {grad}\,(h_i + \Omega _i)%
\eqno(5)
\end{equation*}
Let us note that the motion of each constituent is described by the same
equation that for a single fluid [10,19]. In applications, the motions are
supposed to be isothermal and $\;\theta _i = \theta $ ($\theta $ common
temperature value for all the components). This case corresponds to strong
heat exchange between components.

Eq. (5) yields:
\begin{equation*}
\Gamma _i + \hbox {grad}\, (\varphi_i + \Omega _i) = 0
\end{equation*}
where $\varphi_i = h_i - \theta s_i $ is the free specific enthalpy (or
chemical potential) of constituent i. In the bulks, the volumic free energy
associated with $\alpha $ is denoted by $g_o = g_o (\rho _1, \rho _2,
\theta).$ A complete study of the barycentric motion of the mixture and of
the equation of energy is given in [9]. This generalizes results for single
fluid [3,6,11].

\section{Motion of an isothermal fluid mixture through a plane interface}

An interface in a two-phase mixture is generally schematized by a surface
without thickness. Far from critical conditions, this layer is of molecular
size and density and entropy gradients are very large. A continuous model
schematizes such areas by using an energy in form (4) extending forms given
in [2,16] for compressible fluids.

For the same reasons than for a single fluid, motion is supposed to be
isothermal, stationary and one dimensional with respect to $x$.

The acceleration of component \emph{i} is:
\begin{equation*}
\Gamma _{i}={\frac{1}{2}}\,{\frac{d}{dx}}\,({\frac{q_{i}^{2}}{\rho _{i}^{2}}}%
).
\end{equation*}%
and the equations of motion yield:
\begin{equation*}
\left\{
\begin{array}{l}
\Gamma _{1}=\hbox {grad}\,\{C_{1}\,\Delta \rho _{1}+D\,\Delta \rho
_{2}-g\,_{o,_{\rho _{1}}}\} \\
\\
\Gamma _{2}=\hbox {grad}\,\{D\,\Delta \rho _{1}+C_{2}\,\Delta \rho
_{2}-g\,_{o,_{\rho _{2}}}\}%
\end{array}%
\right.
\end{equation*}%
or:
\begin{equation*}
\left\{
\begin{array}{l}
\displaystyle C_{1}\,\rho _{1}^{\prime \prime }+D\,\rho _{2}^{\prime \prime
}=g\,_{o,_{\rho _{1}}}+{\frac{1}{2}}{\frac{q_{1}^{2}}{\rho _{1}^{2}}}+k_{1}
\\
\\
\displaystyle D\,\rho _{1}^{\prime \prime }+C_{2}\,\rho _{2}^{\prime \prime
}=g\,_{o,_{\rho _{2}}}+{\frac{1}{2}}{\frac{q_{2}^{2}}{\rho _{2}^{2}}}+k_{2}%
\end{array}%
\right.
\end{equation*}

(The derivatives along the motion axis are denoted by $^{\prime }$).

Combination of the last two equations yields the first integral:
\begin{equation*}
g_{o}+k_{1}\,\rho _{1}+k_{2}\,\rho _{2}-{\frac{1}{2}}\,{\frac{q_{1}^{2}}{%
\rho _{1}}}-{\frac{1}{2}}\,{\frac{q_{2}^{2}}{\rho _{2}}}\,-({\frac{1}{2}}%
\,C_{1}\,\rho _{1}^{^{\prime }2}+D\,\rho _{1}^{^{\prime }}\,\rho
_{2}^{^{\prime }}+{\frac{1}{2}}\,C_{2}\,\rho _{1}^{^{\prime }2})=k_{3}
\end{equation*}%
In each bulk, densities have zero-gradients. Eliminating constants $k_{i}$,
dynamical conditions through the interfacial layer yield:
\begin{equation*}
\left\{
\begin{array}{l}
\lbrack g_{,\rho _{1}}(\rho _{1},\rho _{2})]=0 \\
\lbrack g_{,\rho _{2}}(\rho _{1},\rho _{2})]=0\  \\
\lbrack g-\rho _{1}\,g_{,\rho _{1}}-\rho _{2}\, g_{,\rho _{2}}]=0%
\end{array}%
\right. \eqno(6)
\end{equation*}

where [$\;\;$] denotes the discontinuities through the layer and
\begin{equation*}
g=g_{o}-{\frac{1}{2}}\,{\frac{q_{1}^{2}}{\rho _{1}}}-{\frac{1}{2}}\,{\frac{%
q_{2}^{2}}{\rho _{2}}}.
\end{equation*}%
In case of equilibrium ($q_{1}$ and $q_{2}$ are null), the minimum of the
total free energy, with a given total mass for each constituent, yields
conditions (6). By adding the term $\displaystyle-{\frac{1}{2}}\,{\frac{%
q_{1}^{2}}{\rho _{1}}}-{\frac{1}{2}}\,{\frac{q_{2}^{2}}{\rho _{2}}}$ to $%
g_{o}$, the study of flat interfaces crossed by components of a mixture
turns back to an equilibrium problem. In fact, a complete study of the
thickness of the interfacial layer or of spherical interfaces of microscopic
size requires the non-linear model.

 It is more classical to use the
variables $v$ or $c$. The mapping $\;(\rho _{1},\,\rho _{2},\,g)\
\rightarrow \ (v,c,G)\;$ where $v=\displaystyle{\frac{1}{\rho _{1}+\rho _{2}}%
}\,,\,c={\frac{\rho _{2}}{\rho }}\,,\,\displaystyle G_{o}={\frac{g_{o}}{\rho
}}$ and $G=G_{o}-\displaystyle{\frac{1}{2}}\,v^{2}\,({\frac{q_{1}^{2}}{1-c}}+%
{\frac{q_{2}^{2}}{c}})\;$, allows to write
\begin{equation*}
\left\{
\begin{array}{l}
\lbrack G_{,v}(v,c)]=0 \\
\lbrack G_{,c}(v,c)]=0 \\
\lbrack G-v\,G_{,v}-c\,G_{,c}]=0%
\end{array}%
\right. \eqno(7)
\end{equation*}
Conditions (6) or (7) express that the points corresponding to the bulks of
the Gibbs surface $(\Sigma )$ associated with the dynamical free energy $g$
or $G$ are contact points of a bitangent plane (Fig. 2). 

\begin{figure}[h]
\begin{center}
\includegraphics[width=11cm]{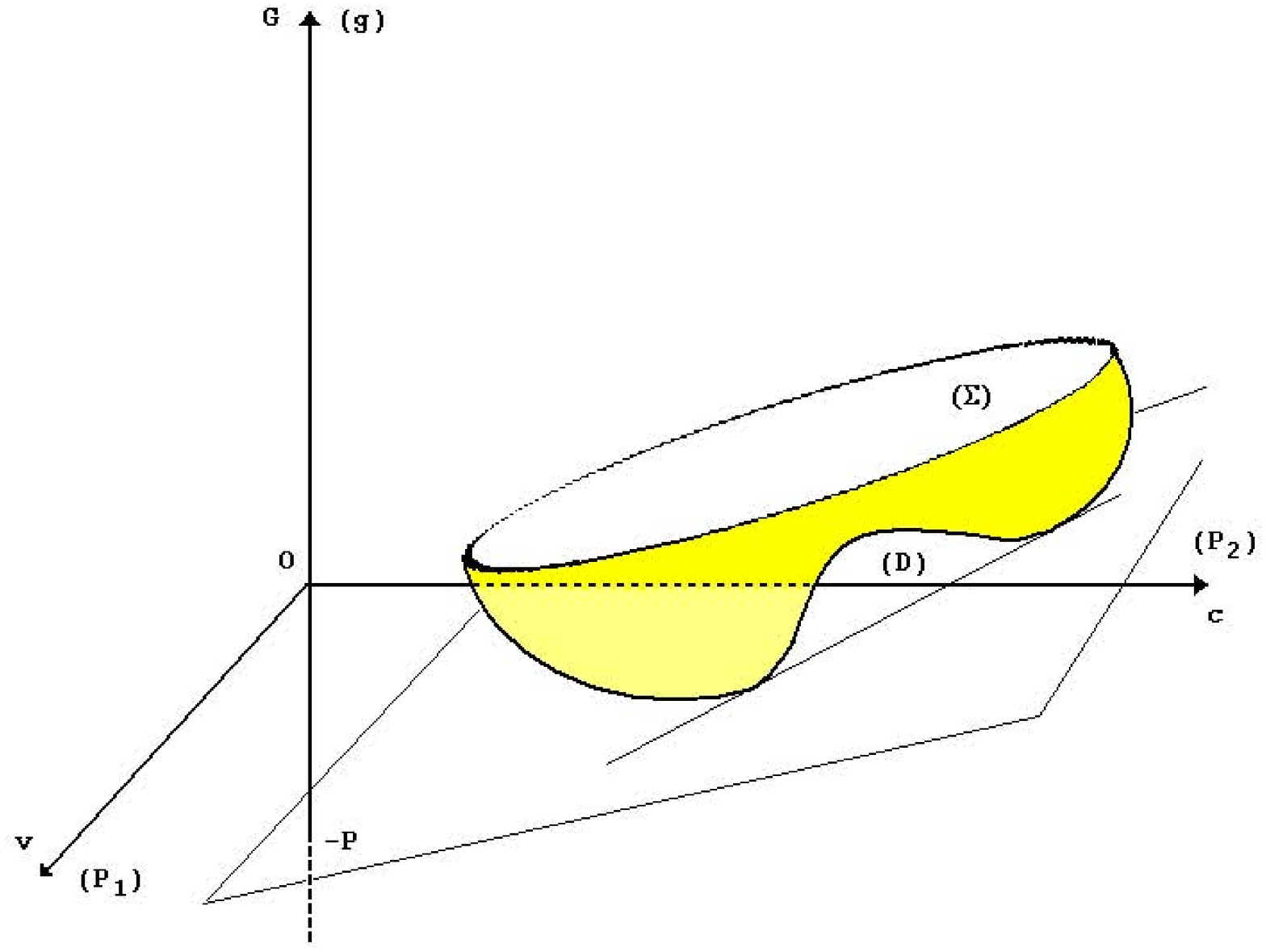}
\end{center}
\end{figure}

\vskip 0.25cm \centerline{Figure 2}

The pressure of the mixture is $p = - G_{o, v} (v, c)$, the dynamical
pressure is $P = - G_ {,v} (v, c)$. \newline
The $z$-coordinate of the point of intersection of the bitangent plane with
the $g$-axis is $\;-P\;$. The chemical potential $G_{,c}\;$ is denoted by $%
\Phi $.

Contact points of the bitangent plane with surface $(\Sigma)$ generate a
curve. Along this curve, the straight line ($D$) connecting the contact
points is the \emph{characteristic line} of the plane. Along the line, the
relation
\begin{equation*}
(v - v_0)\, d \,G_{, v} + (c - c_0)\, d \, G_{,c} = 0
\end{equation*}
allows to generalize Clapeyron formula such that:
\begin{equation*}
{\frac{dP }{d\Phi}} = {\frac{c_1 - c_0 }{v_1 - v_0}}
\end{equation*}
where subscripts 0 and 1 are associated with the two bulks.

\section{Maxwell rules for fluid mixtures}

At a given temperature $\theta $, the differential form $dG = -P\, dv +
\Phi\, dc\ $ yields the relation:
\begin{equation*}
{\int_{(C)}} \,(\Phi - \Phi_0) \,dc - (P - P_0)\, dv = 0
\end{equation*}
where $(C)$ is an arbitrary curve connecting the two bulks in the space $%
(c,v)$. The common values of $\Phi$ and $P$ in the bulks are denoted by $%
\Phi_0$ and $P_0$.

We can calculate the integral along special paths:

Along the curve $(C)$ given by the implicit function $c (v)$ defined such
that $G_ {,c} \,(c (v), v) = \Phi_0$, the relation
\begin{equation*}
{\int_{(C)}} \,(P - P_0) \,d\mathit{v} = 0
\end{equation*}
represents the Mawxell equal area rule for the path {$\mathbf{\overline {a b
c d} }$} in the plan $(P,v)$ where\footnote{$(C)$ is not the curve associated with $P = P(c,v)$ at $c$ constant.} $P =
-G_{, v}\, (c\,(v), v)$

Likely, for $P = P_0$, the relation:
\begin{equation*}
\int_{(C)}\ (\Phi - \Phi_0) \,dc =0
\end{equation*}
represents the Mawxell equal area rule for the path {$\mathbf{\overline {a b
c d} }$} in the plan $(c,\Phi)$ where $\Phi = G_{,c}\,(c,v(c))$ is such that
$v$ is expressed as a function of $c $ by the relation $G_{,v}\, (c,v (c)) =
-P_0.$

\begin{figure}[h]
\begin{center}
\includegraphics[width=11cm]{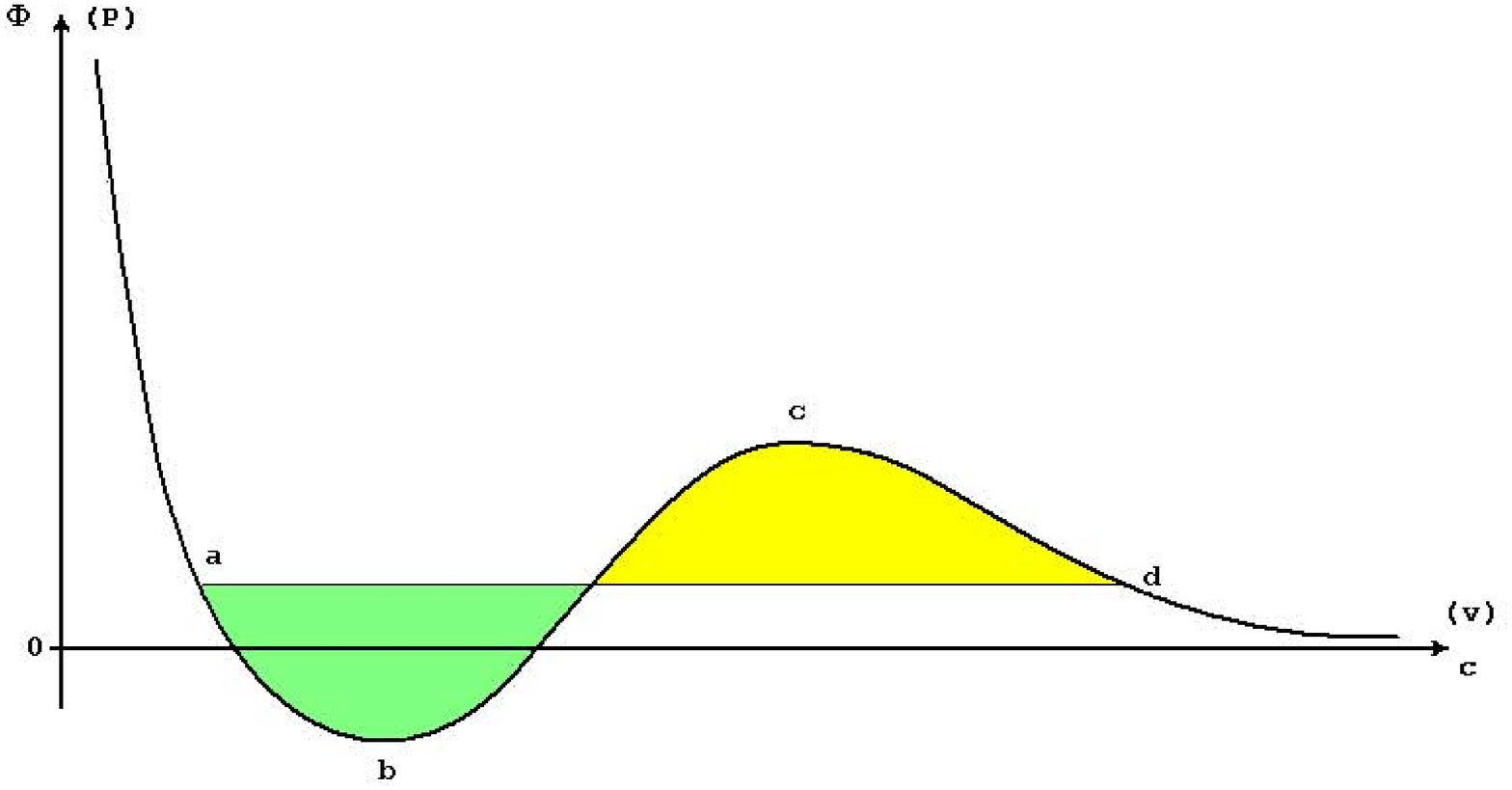}
\end{center}
\end{figure}

\vskip 0.25cm

\centerline{Figure 3}

These results generalize the ones obtained by Slemrod in the case of
inviscid flows of single fluids [21]. The knowledge of the free specific
energy $G_o$ is sufficient to solve the previous problem. Function $G_o$ is
obtained by integration of the equation of state $p = p\, (v, c, \theta)$:
\begin{equation*}
{\frac{\partial\,G_o\, (v, c, \theta) }{\partial v}} = -p\, ( v, c, \theta) %
\eqno(8)
\end{equation*}
Function $p$ is determined except for an additional function of $c$ and $%
\theta $. With the physical assumption that $G_o$ is independent of $c$ when
$v$ vanishes, the additional function depends only on $\theta $.

The simplest model of equation of state is by van der Waals:
\begin{equation*}
p = {\frac{R \theta }{v - b}}\,-\,{\frac{a }{v^2}}\eqno(9)
\end{equation*}
where $a$ and $b$ are given by the mixing rules:
\begin{equation*}
a^2 = (1 - c)\, a^2_1 + c\, a^2_2\ \ \hbox{et}\ \ b = (1 - c) b_1 + c\, b_2 %
\eqno(10)
\end{equation*}
(Subscripts 1 and 2 are associated with the components 1 and 2).

One deduces: $\displaystyle G_o = -R \,\theta \, Ln\, ( v - b) - {\frac{a }{v%
}}.$

Other equations of state are proposed for example $\displaystyle p = {\frac{%
R\, \theta }{v - b}} - {\frac{a }{v ( v + b)}}$ (particularly in petroleum
industry [5,15,18]) or by changing the form of the mixing rules (other
expressions than $\displaystyle
a^2 = \sum_{i=1}^{n}\, a^2_i\, c_i\ \ \hbox{and}\ \ b =
\sum_{i=1}^{n}\,b_i\, c_i$ can be considered).

Equations (7), (8), (9) and (10) lead to the system:
\begin{equation*}
\left\{
\begin{array}{l}
\displaystyle\lbrack \;{\frac{\theta }{v-b}}+{\frac{v}{1-c}}\ q_{1}^{2}+{%
\frac{v}{c}}\ q_{2}^{2}-{\frac{(b+1)^{2}}{v^{2}}}\;]=0 \\
\displaystyle\lbrack \;\theta \{1-Ln\,(v-b)+{\frac{b_{1}}{v-b}}\}+{\frac{1}{2%
}}\ {\frac{v^{2}}{(1-c)^{2}}}q_{1}^{2}-2{\frac{(b_{1}+1)\,(b+1)}{v}}\;]=0 \\
\displaystyle\lbrack \;\theta \{1-Ln\,(v-b)+{\frac{b_{2}}{v-b}}\}+{\frac{1}{2%
}}\ {\frac{v^{2}}{c^{2}}}q_{2}^{2}-2{\frac{(b_{2}+1)\,(b+1)}{v}}\;]=0%
\end{array}%
\right.
\end{equation*}

which is a linear system with respect to $\theta ,q_{1}^{2},q_{2}^{2}$.

\section{Conclusion}

The study of fluid mixture motions by a model taking into account density
gradients of components requires the knowledge of the global free energy. In
fact, only physical experiments and molecular theories provide an equation
of state (in van der Waals type). Isothermal motions can be studied by using
such an equation and yield the jump conditions through interfacial layers.
To carry the program one step further and to study non-isothermal motions,
it is necessary to get additional knowledges (such that specific heats...).

\textbf{Acknowledgements}: \textsl{I am grateful to Professor P. Casal for
his help and to the Institute of Mathematics and its Applications for
hospitality and financial support in the month of October 1990. This
research was supported in part by the Institute for Mathematics and its
Applications with funds provided by the National Science Foundation and in
part by the French Foreign Office.}

\end{document}